\documentclass[aps,pra,twocolumn,groupedaddress,floatfix]{revtex4}
\usepackage{bm} \usepackage{graphicx}
\usepackage{epsfig}
\usepackage{subeqn}

\begin{document}


\title{Bogoliubov modes of a dipolar condensate in a cylindrical trap}

\author{Shai Ronen} 
\affiliation{JILA and Department of Physics, University of Colorado, Boulder, CO 80309-0440}
\author{Daniele C. E. Bortolotti} 
\affiliation{JILA and Department of Physics, University of Colorado, Boulder, CO
80309-0440} 
\affiliation{LENS and Dipartimento di Fisica, Universit\'{a} di Firenze, Sesto Fiorentino, Italy}
\author{John L. Bohn} 
\affiliation{JILA, NIST, and Department of Physics, University of Colorado,
 Boulder, CO 80309-0440 }
\email{bohn@murphy.colorado.edu}

\affiliation{}


\date{\today}

\begin{abstract}

The calculation of properties of Bose-Einstein condensates with
dipolar interactions has proven a computationally intensive problem
due to the long range nature of the interactions, limiting the scope
of applications. In particular, the lowest lying Bogoliubov
excitations in three dimensional harmonic trap with cylindrical
symmetry were so far computed in an indirect way, by Fourier analysis
of time dependent perturbations, or by approximate variational
methods. We have developed a very fast and accurate numerical
algorithm based on the Hankel transform for calculating properties of
dipolar Bose-Einstein condensates in cylindrically symmetric traps.
As an application, we are able to compute many excitation modes by
directly solving the Bogoliubov-De Gennes equations. We explore the
behavior of the excited modes in different trap geometries. We use
these results to calculate the quantum depletion of the condensate by
a combination of a computation of the exact modes and the use of a
local density approximation.

\end{abstract}


\maketitle

\section{Introduction \label{intro}}

The realization of a Bose-Einstein condensate (BEC) of $^{52}$Cr
\cite{Stuhler05} marks a major development in degenerate quantum gases
in that the inter-particle interaction via magnetic dipoles in this
BEC is much larger than those in  alkali atoms, and leads to an
observable change in the shape of the condensate. The long range
nature and anisotropy of the dipolar interaction pose challenging
questions about the stability of the BEC and brings about unique
phenomena, such as roton-maxon spectrum and different phases of vortex
lattices. They also present a significant theoretical challenge,
especially for the calculation of excitations
\cite{Yi00,Goral00,Yi01,Yi02,Baranov02,Goral02,Santos03,ODell04,Nho05,Cooper05}.

For $N$ bosons in an external trap potential $V_{ext}(\bm{r})$ at very
low temperatures, the condensate can be described using mean-field
theory \cite{Bortolotti06,Ronen06}. All the particles in the condensate then have the same wave
function $\Psi(\bm{r})$, which is described by the following
time-dependent Gross-Pitaevskii equation (GPE):
\begin{eqnarray}
\lefteqn{i\frac{\partial\Psi(\bm{r},t)}{\partial t}
=\Bigg[-\frac{\hbar^{2}}{2m}\nabla^{2}+U(\bm{r})+} \nonumber \\
& &(N-1)\int d\bm{r'}
V(\bm{r}-\bm{r'}) |\Psi(\bm{r'},t)|^2\Bigg]\Psi(\bm{r},t),
\label{gpe}
\end{eqnarray}
where $\bm{r}$ is the displacement from the trap center, $m$ is the
atomic mass, and the function $\Psi$ is normalized to unit norm. We
shall consider the case of a cylindrical harmonic trap, with
$U(\bm{r})=\frac{1}{2}(\omega_{\rho}^{2}
\rho^{2}+\omega_{z}^{2}z^{2})$. For dipolar interactions the
potential $V(\bm{r})$ may be written\cite{Yi00,Yi01}:
\begin{eqnarray}
V(\bm{r})=\frac{4\pi\hbar^{2}a}{m}\delta(\bm{r}) + d^2\frac{1-3\cos^{2}\theta}{r^3},
\label{pseudo}
\end{eqnarray}
where $a$ is the scattering length, $d$ the dipole moment, $\bm{r}$
the distance between the dipoles, and $\theta$ the angle between the
vector $\bm{r}$ and the dipole axis, which we shall take to be aligned
along the trap axis. Note that in general $a$ depends on the dipole
moment \cite{Yi00,Yi01,Bortolotti06,Ronen06}. This observation is
important in an experimental setup where the dipole moment is tuned by
an external field. In the present work we shall rather assume that $a$
and $d$ are either fixed or may be tuned independently.

For the following we define the transverse harmonic oscillator length
$a_{ho}=\sqrt{\frac{\hbar} {m \omega_{\rho}}}$, and the two
dimensionless interaction parameters $s=(N-1)\frac{a}{a_{ho}}$ and
$D_{*}=(N-1)\frac{m}{\hbar^2}\frac{d^2}{a_{ho}}$. We work in scaled
units where $\hbar=m=1$.

Eq.~(\ref{gpe}) with the potential Eq.~(\ref{pseudo}) is an
integro-differential equation. The integral term $d^2\int
d\bm{r'}\frac{1-3\cos^{2}\theta}{|\bm{r}-\bm{r'}|^{3}}
|\Psi(\bm{r'})|^{2}$ needs special attention due to the apparent
divergence of the dipolar potential at small distances. Moreover, it
is non-local and requires a computationally expensive 3-dimensional
convolution. Two different procedures \cite{Goral02, Yi01} have been
proposed to deal with this problem. In Ref.~\cite{Goral02}, Fourier
transformation to momentum space was employed to perform the
convolution and avoid the divergence, but the calculation still had to
be performed in 3D even for cylindrical symmetry. In Ref.~\cite{Yi01} the
cylindrical symmetry of the system was utilized to compute a 2D
analytic expression for the convolution kernel in real space, which,
however, required further computationally intensive numerical
smoothing in order to avoid the $r=0$ singularity. In either case, it
seems that the computations remain quite intensive, and this may be
the reason why calculations of low lying excitation modes
\cite{Yi02,Goral02} were performed only via spectral analysis of time
dependent perturbations. These calculations were time consuming and
numerically challenging. 

In this paper, we present a new algorithm based on Hankel transform
for calculating properties of dipolar condensates in cylindrical traps
which combines the advantages of the two methods above: namely, both a
transform to momentum space and utilization of the cylindrical
symmetry to work in two dimensions. In addition, we examine the
accuracy of the previous 3D Fourier transform algorithm and show that
it does not achieve high spectral accuracy as might have been
expected. We analyze the reason for this and suggests a simple
correction that provides high accuracy. The same correction also
applies to our new 2D algorithm.

The ground state is found by minimizing the total energy, which is
often done by propagating the wavefunction in imaginary time. We find
that this method is slow. Instead, we use a highly efficient
minimization technique, conjugate-gradients, to further speed up the
calculation of the ground state, resulting in a typical computation
time of fraction of a second on a present day PC. Finally, we apply
our new algorithm to a fast and efficient calculation of many excited
modes via direct solution of the Bogoliubov-De Gennes (BdG) equations. Calculation of tens of
modes is achieved in a span  of a few seconds to a few
minutes. We also compute the quantum depletion at $T=0$ due to the
dipolar interaction.

\section{The algorithm}

Following \cite{Goral02}, the calculation of the dipolar interaction
integral can be simplified by means of the convolution theorem. Let
$n(\bm{r})=|\Psi(\bm{r})|^2$ be the density per particle at $\bm{r}$
and let $V_{D}(\bm{r})= (3 z^{2}/r^{2}-1)/r^3$. Let
$\tilde{n}(\bm{k})$ and $\tilde{V}_{D}(\bm{k})$ be their Fourier
transforms, i.e
\begin{eqnarray}
\tilde{n}(\bm{k})=\int d\bm{r} \exp(- i \bm{k} \cdot \bm{r})n(\bm{r}),
\end{eqnarray}
etc. Then the mean field $\Phi_{D}(\bm{r})$ at the point $\bm{r}$ due
to the dipolar interaction is:

\begin{eqnarray}
\Phi_{D}(\bm{r})=\int d\bm{r'} V_{D}(\bm{r}-\bm{r'}) n(\bm{r'}) = \mathcal{F}^{-1} \left( \tilde{V}_{D}(\bm{k})n(\bm{k}) \right),
\label{dipintegral}
\end{eqnarray}
where $\mathcal{F}^{-1}$ is inverse Fourier transform. The Fourier
transform of $V_{D}$ may be found by expanding
$\exp(i\bm{k} \cdot \bm{r})$ in a series of spherical harmonics and
spherical Bessel functions (the usual expansion of a free planar wave
in free spherical waves), where only the $Y_{20}$ term gives non-zero
contribution. The result is \cite{Goral02}:

\begin{eqnarray}
\tilde{V}_{D}(\bm{k})=\frac{4\pi}{3}
\left( 3 k_{z}^{2}/k^{2}-1 \right) = \frac{4\pi}{3}\left( 3\cos^{2}\alpha-1 \right),
\label{dippot}
\end{eqnarray}
where $\alpha$ is the angle between $\bm{k}$ and the $z$ axis.

It has also been shown in Ref.~\cite{Goral02} that a cutoff of the dipolar
interaction at small $\bm{r}$ in real space is not important when the
calculation is performed in momentum space, since it affects only very
high momenta which are not sampled. $n(\bm{k})$ can be numerically
evaluated from $n(\bm{r})$ by means of a standard fast Fourier
transform (FFT) algorithm. This method is quite general, and has the
speed advantage of using FFT. It is similar to the calculation of the
kinetic energy to high accuracy by Fourier transform of the
wavefunction to momentum space and multiplication by $-k^{2}/2$.

We wish to consider traps with cylindrical symmetry, for which the the
projection of the angular momentum on the $z$-axis is a conserved
quantity. The eigenstates may then be written
\begin{eqnarray}
\psi(\rho,\phi,z)=\exp(i m \phi)G(\rho,z),
\label{cyl}
\end{eqnarray}
with $(\rho,z,\phi)$ the usual cylindrical coordinates, and $m$ an
integer. For the ground state $m$ is zero, but we would also like to consider
$m>0$ excitations.

The 3D FFT cannot take direct advantage of the cylindrical
symmetry. The key idea of the new algorithm is the observation that
for a function of the form (\ref{cyl}), the 2D Fourier transform in the
$(x,y)$ plane is reduced, after integration over $\phi$, to a 1D
Hankel (or Bessel) transform of order $m$ \cite{Arfken}:

\begin{eqnarray}
\tilde{\psi}(k_{\rho},k_{\phi},z)=2 \pi i^{-m} e^{ i m k_{\phi}} \int_{0}^{\infty} G(\rho,z) J_{m}(k_{\rho}\rho) \rho d\rho,
\end{eqnarray}
where $J_{m}(x)$ is the Bessel function of order $m$. Thus, a
combination of Hankel transform in the transverse ($\rho$) direction
and Fourier transform in the axial $(z)$ direction, which we shall
call here the discrete Hankel-Fourier transform (DHFT), can be used to move
between spatial and momentum spaces. The wavefunction need only be
specified on a two dimensional grid in $(\rho,z)$ coordinates. The
transform can be used to compute both the dipolar interaction energy and the
kinetic energy. 

Interestingly, there exist fast Hankel transforms \cite{Siegman77,Talman78,Hamilton00}
\footnote{Fast Hankel transform code available from
http://casa.colorado.edu/~ajsh/FFTLog}. These algorithms perform a
discrete transform of $N$ radial points in time complexity of $O(N
\log(N))$, just as for 1D fast Fourier transform (we use $N$ also to
indicate number of particles. The meaning should be clear from the
context). However, they require sampling the transformed function on a
logarithmically spaced grid. Thus the function is oversampled at small
radii. Our experience with the fast algorithm \cite{Talman78} is that
it works well for one transform, but requires careful attention and
tuning to avoid increasing numerical errors when applied many times
over in the energy minimization process (for an early attempt of
applying the method \cite{Siegman77} for a scattering problem, see
\cite{Bisseling85}). For this reason, We have selected as our method
of choice a discrete Hankel transform (DHT) with grid sampling based
on zeros of the Bessel function
\cite{Yu98,Guizar04,Lemoine94,Lemoine97,Lemoine03}
\footnote{Matlab code available from
www.mathworks.com/matlabcentral/fileexchange as ``Integer order Hankel
transform''}. We found it to be accurate and robust. This algorithm is
of complexity $O(N^2)$. The non-uniform sampling might seem unusual at
first, but is actually of great advantage in that, as we shall show,
it allows a highly accurate integration formula, reminiscent of
Gaussian-quadrature. Although for large grids the DHT is slower than
the fast Hankel transforms mentioned above, we find that in practice
it is faster for radial grid sampling of up to 32 points.

\subsection{The discrete Hankel transform}

We shall now describe the DHT. The Hankel transform of order $m$ of a
function $f(r)$ is defined by:

\begin{eqnarray}
\tilde{f}(k)= \int_{0}^{\infty} f(\rho) J_{m}(k \rho) \rho d\rho.
\label{hankeltrans}
\end{eqnarray}
The inverse transform is obtained by simply exchanging $f$ and
$\tilde{f}$ along with $\rho$ and $k$.

Let us assume that the function $f(r)$ is practically zero for $r \geq
R$, and $\tilde{f}(k)$ is practically zero for $k \geq K$. Let $g(\rho)$
be sampled at $N$ points
\begin{eqnarray}
\rho_{j}=\alpha_{mj}/K,\ j=1,\ldots,N 
\label{rindex}
\end{eqnarray}
where $\alpha_{mj}$ is the $j$'th root of $J_{m}(\rho)$. Let $\tilde{f}(k)$ be sampled at $N$ points 
\begin{eqnarray}
k_{i}=\alpha_{mi}/R,\ i=1,\ldots,N.
\label{kindex}
\end{eqnarray}
Then Eq.~(\ref{hankeltrans}) is approximated \cite{Lemoine94} by the discrete sum:
\begin{eqnarray}
\tilde{f}(k_i)= \frac{2}{K^2}\sum_{j=1}^{N} \frac{f(\rho_j)}{J_{m+1}^2(\alpha_{mj})} J_{m}(\frac{\alpha_{mj}\alpha_{mi}}{S}), 
\label{quasihankel}
\end{eqnarray}
where $S=RK$. 
Eq.~(\ref{quasihankel}) can be written in a more convenient form by defining 
\begin{eqnarray*}
F(j)&=&\frac{R}{|J_{m+1}(\alpha_{mj})|}f(\alpha_{mj}/K) ,\\
\tilde{F}(i) &=& \frac{K}{|J_{m+1}(\alpha_{mi})|} \tilde{f} (\alpha_{mi}/R) ;
\end{eqnarray*}
thus Eq.~(\ref{quasihankel}) reduces to
\begin{eqnarray}
\tilde{F}(i)=\sum_{j=1}^{N}T_{ij}F(j),
\end{eqnarray}
where
\begin{eqnarray}
T_{ij}=\frac{2 J_{m}(\alpha_{mi}\alpha_{mj}/S)}{|J_{m+1}(\alpha_{mi})||J_{m+1}(\alpha_{mj})|S}
\end{eqnarray}
defines the elements of an $N$x$N$ transformation matrix $\bm{T}$.

$\bm{T}$ is a real, $N$x$N$ symmetric matrix. Note also that it depends
on $S$. Imposing the boundary conditions $f(R)=\tilde{f}(K)=0$
requires $S=\alpha_{m,N+1}$ (then $T_{(N+1),j)}=0$). Since the Hankel
transform is the inverse of itself, we should require $\bm{T}$ to be
unitary for self-consistency. In fact, with $S=\alpha_{m,N+1}$, $\bm{T}$ is found to be very close to
being unitary \cite{Guizar04}. E.g., for $N=50$ and $m=0-3$, $|\det[\bm{T}]|-1<10^{-8}$, and
the unitarity is better with larger $N$. If an exactly unitary matrix $\bm{T}$
is desired, it has been suggested \cite{Lemoine03} using
$\bm{B}=(\bm{T}^{\dag}\bm{T}\bm{T})^{-1/2}\bm{T}$ instead of $\bm{T}$. But in
numerical tests it was found that in practice essentially the same
high accuracy was obtained with $\bm{T}$ as with $\bm{B}$.

In practice, we first determine an appropriate $R$ for the function, and
some convenient $N$. Then define $S=\alpha_{m,N+1}$, and $K=S/R$. $N$
should be chosen such that $\tilde{f}(k)$ is as small as desired for
$k \geq K$.

\subsection{Quadrature-like integration formula}

When the integrals in Eq.~(\ref{gpe}) are evaluated in cylindrical
coordinates, we are required to calculate an integral of the form:
\begin{eqnarray}
I[f]=\int_{0}^{\infty} f(\rho) \rho d\rho 
\label{integral1}
\end{eqnarray}

An important benefit of sampling the function $f(\rho)$ according to
Eq.~(\ref{rindex}) is that we are able to derive a highly accurate
approximation to this integral. Let $\tilde{f}(k)$ be the $m$'th order
Hankel transform of $f(\rho)$, Eq.~(\ref{hankeltrans}). We shall fist
assume that $f(\rho)$ is band limited. That is, we assume that there exist $K$ such that
\begin{eqnarray}
\tilde{f}(k)=0 \; (k>K). 
\label{eq:condition}
\end{eqnarray}
Then $I[f]$ is given \textit{exactly} by the following series:
\begin{eqnarray}
I[f]=\frac{2}{K^2}\sum_{i=1}^{\infty}\frac{1}{J_{m+1}^{2}(\alpha_{mi})}f(\alpha_{mi}/K).
\label{integral2}
\end{eqnarray}
We give a simple proof of this formula. First, note that by
Eq.~(\ref{hankeltrans}), $I[f]=\tilde{f}(0)$. Expand $\tilde{f}(k)$ on
$[0,K]$ in a Fourier-Bessel series \cite{Arfken} to find:
\begin{eqnarray}
\tilde{f}(k)=\frac{2}{K^2}\sum_{i=1}^{\infty}\frac{1}{J_{m+1}(\alpha_{mi})^2 } f(\alpha_{mi}/K)
J_{m}(\alpha_{mi}k/K).
\end{eqnarray}
Substitution of $k=0$ gives immediately Eq.~(\ref{integral2}). This
formula has appeared relatively recently in the computational
mathematics literature \cite{Frappier93,Grozev95,Ogata05}. It has been
shown to be intimately related to Gaussian quadrature. In Gaussian
quadrature, the sampling points are roots of orthogonal polynomials,
while here they are roots of the Bessel function $J_{m}(K \rho)$.

The wavefunctions of interest in our problem are not strictly zero for
$k>K$, but they typically decrease exponentially for large $k$.
Moreover, as the wavefunction decreases exponentially in space,
the infinite series can be truncated to provide the following
approximate formula:
\begin{eqnarray}
I[f]= \frac{2}{K^2}\sum_{i=1}^{N}\frac{1}{J_{m+1}(\alpha_{mi})^2} f(\alpha_{mi}/K).
\end{eqnarray}
In our application, $N$ and $K$ are the same as those given above for the DHT. The approximation converges
exponentially to the exact value with increasing $N$ and $R$.

\subsection{Accuracy of calculating the dipolar interaction energy}

Before continuing to applications, we re-examine the accuracy of the
3D FFT method \cite{Goral02}. The behavior discussed below appears
also in the 2D method, but the analysis in the 3D case is easier. From
the similarity to the calculation of kinetic energy with spectral
accuracy, which typically achieves machine precision with a small
number of grid points, we expected that the dipolar interaction energy
will also be calculated to this high accuracy. We find that the
relative accuracy, typically varying between 1E-5 to 1E-2, is enough
for practical purposes, but not as accurate as might be expected. A
detailed analysis is given in the appendix. The reason for the
numerical errors is traced down to the discontinuity of
$\tilde{V}_D(\bm{k})$, Eq.~(\ref{dippot}), at the origin. An improved
accuracy, by at least two orders of magnitude and up to machine
precision, is obtained by using, instead of Eq.~(\ref{dippot}), the
Fourier transform of a a dipolar interaction truncated to zero outside a
sphere of radius $R$, where the spatial grid dimensions are
$[-R,R] \times [-R,R] \times [-R,R]$. Thus, Eq.~(\ref{eq:cutR}) replaces
Eq.~(\ref{dippot}). Note that this truncation of the dipolar potential
has no physical effect on the system, as long as $R$ is greater than
the condensate size. For a pancake trap it is advantageous to use a grid
of dimensions $[-P,P] \times [-P,P] \times [-Z,Z]$ with $Z<P$. For $Z<P/2$ we
find that it is preferable to use the Fourier transform of dipolar
interaction truncated to zero for $|z|)>Z$, Eq.~(\ref{eq:cutZ}). These
modifications apply also to the 2D case.

\section{Ground state of a dipolar condensate \label{sec:ground}}

For finding the ground state of dipolar BEC, the wavefunction is
sampled on a 2D grid $(\rho_{i},z_{j})$, with $\rho_{i}$
determined from Eq.~(\ref{rindex}), and $z_{j}$ evenly sampled. Since the
ground state is symmetric in $z$, it is enough to sample $z \geq
0$. The FFT in the $z$ direction is then performed as a fast cosine
transform \cite{NumericalRecipes}, for which we used the FFTW software
package \footnote{http://www.fftw.org}. The fast cosine transform uses
the property of the wavefunction being real and symmetric to enhance
speed by a factor of 4 compared to standard complex FFT.

The most commonly practiced method to obtain the ground state is
propagation of an initial guess of the wavefunction in imaginary time,
using Eq.~(\ref{gpe}) with $t \rightarrow -i t$. This method is
robust but slow, though the reduction of the problem to 2D speeds it
up considerably. We obtained a further substantial gain in speed by
adopting, instead, direct minimization of the total energy using the
conjugate-gradients technique \cite{NumericalRecipes}. This technique
has become popular in density functional theory calculations, and a
review of it appears in Ref.~\cite{Payne92}. A previous application to BEC
vortices is found in Ref.~\cite{Modugno03}, which we closely follow. The GP
energy functional is given by
\begin{eqnarray}
\lefteqn{E[ \Psi , \Psi^{*}] = \int d\bm{r} \Psi^{*}(\bm{r}) H_{0} \Psi(\bm{r})
 + \nonumber} \\
& &\frac{N-1}{2}\int \int d\bm{r} d\bm{r'}
 |\Psi^{2}(\bm{r'})|V(\bm{r}-\bm{r'}) |\Psi^{2}(\bm{r})|,
\label{eq:func1}
\end{eqnarray}
where the Hamiltonian operator $H_0$ contains the kinetic and the trap potential
terms
\begin{eqnarray}
H_0=-\frac{\hbar^2\nabla^2}{2m}+U(\bm{r}),
\end{eqnarray}
and the condensate wave function $\Psi$ obeys the normalization condition
\begin{eqnarray}
\| \Psi \| \equiv \int d\bm{r}|\Psi(\bm{r})|^2=1.
\label{eq:constraint}
\end{eqnarray}

The essence of the conjugate-gradients method is minimizing the
energy, Eq.~(\ref{eq:func1}) by successive
line-minimizations along optimally chosen directions.
The initial direction is along the gradient, but in the $(n+1)$'th step the new direction
is a judicious linear combination of the new gradient and the previous
($n$'th) direction. This 'memory' property provides the algorithm
with a better feeling (so to speak) for the shape of the energy
surface, and faster convergence is achieved compared to simply
following the present gradient at each step.
An important feature for our specific problem is that that each
line-minimization step can be done analytically. Another
important point is the need to ensure that, just like in imaginary
time propagation, we reach the local minimum nearest to the initial
guess, rather than a global minimum, which may be a collapsed
state. Details of the implementation are given in the appendix.

The algorithm using the DHFT and the conjugate-gradients minimization
was implemented in Matlab, and its correctness verified by comparing
to the published results in the literature and to an independent code
implementing the 3D method of \cite{Goral02} with imaginary time
propagation. In a typical application, the grid consists of 32x32
points, covering the domain $[0,8] \times [0,8]$ in the $(\rho,z)$
coordinates. The starting guess is a sufficiently wide Gaussian. We
have checked the numerical convergence of the ground state energy with
respect to increasing the grid resolution and its size. Generally,
convergence will depend on the parameters of the problem, but in most
cases this grid already achieves convergence to very high accuracy.
As a benchmark, the harmonic oscillator energy in a spherical trap
($\omega_{\rho}=\omega_{z}=1$) is obtained to accuracy of $10^{-14}$
with the above grid. For a dipolar BEC with the interaction
parameters $s=1$ and $D_{*}=3$, we find that the energy is
converged to the same accuracy, $10^{-14}$, with respect to increasing
the resolution and size of the grid. The runtime for this computation
on our PC is 0.5 seconds.

\section{Bogoliubov-De Gennes excitations \label{sec:excitations}}
\subsection{Formulation}

We now turn our attention to computing excitations of the condensate
by direct solution of the BdG equations (see, e.g, \cite{BEC2003}, for
the case of a short range potential). We first derive the BdG
equations for the dipolar case by analyzing the linear stability of
the time-dependent GPE about a stationary state $\Psi_0(\bm{r})$. We
write
\begin{eqnarray}
\Psi(\bm{r},t)=[\Psi_{0}(\bm{r})+\vartheta(\bm{r},t)]e^{-i \mu t},
\label{eq:perturb}
\end{eqnarray}
where $\mu$ is the chemical potential of the stationary
state, and $\vartheta$ is a small quantity for which we look for
a solution of the form:
\begin{eqnarray}
\vartheta(\bm{r},t)=\lambda (u(\bm{r})e^{-i \omega t}+\upsilon^{*}(\bm{r})e^{i \omega t}),
\label{eq:form}
\end{eqnarray}
where $\omega$ is the frequency of the oscillation, $\lambda$ the
amplitude of the perturbation ($\lambda<<1$), and $u$ and
$\upsilon$ are normalized according to:
\begin{eqnarray}
\int d \bm{r} [u^2(\bm{r})-\upsilon^2(\bm{r})]=1.
\label{eq:normalization}
\end{eqnarray}

By collecting the terms linear in $\lambda$ and evolving in time like $e^{-i \omega t}$ and
$e^{i \omega t}$, one obtains the following pair of BdG equations:
\begin{eqnarray}
\lefteqn{\omega u(\bm{r}) = } \nonumber\\ 
& & [H_{0}-\mu+ (N-1)\int d\bm{r'}
 \Psi_0^{*}(\bm{r'})V(\bm{r}-\bm{r'})\Psi_0(\bm{r}')]u(\bm{r})+
\nonumber \\
& & (N-1)\int
 d\bm{r'}\Psi_0^*(\bm{r'})V(\bm{r}-\bm{r'})u(\bm{r'})
\Psi_0(\bm{r})+ \nonumber \\
& &(N-1)\int
d\bm{r'}\Psi_0(\bm{r'})V(\bm{r}-\bm{r'})\upsilon(\bm{r'})\Psi_0(\bm{r}),
\label{eq:BdGpair}\\
\lefteqn{-\omega \upsilon(\bm{r}) = } \nonumber \\ 
& & [H_{0}-\mu+ (N-1) \int d \bm{r'}
 \Psi_{0}^{*} (\bm{r'})
 V(\bm{r}-\bm{r'})\Psi_0(\bm{r}')]\upsilon(\bm{r})+ \nonumber \\
& & (N-1)\int d\bm{r'}\Psi_0^*(\bm{r'})V(\bm{r}-\bm{r'})\upsilon(\bm{r'})
 \Psi_0(\bm{r})+ \nonumber \\
& & (N-1)\int
d\bm{r'}\Psi_0(\bm{r'})V(\bm{r}-\bm{r'})u(\bm{r'})\Psi_0(\bm{r}),
\nonumber 
\end{eqnarray}
where $H_0=-\frac{1}{2}\nabla^2+U(\bm{r})$, and $V(\bm{r})$ is given by
Eq.~(\ref{pseudo}). This linear system may be expressed more succinctly by the
matrix form
\begin{widetext}
\begin{eqnarray}
\left(
\begin{array}{cc}
H_0-\mu+C+X & X \\
-X & -H_0+\mu-C-X \\ 
\end{array} \right) \left( \begin{array}{c} u \\ 
\upsilon \end{array} \right) = \omega \left( \begin{array}{c} u
\\ 
\upsilon \end{array} \right)
\label{eq:mat1}
\end{eqnarray}
\end{widetext}
with 
\begin{eqnarray}
\lefteqn{(C \chi)(\bm{r}) = } \nonumber \\
& & (N-1)\int d\bm{r'}
 \Psi_0(\bm{r'}) V(\bm{r}-\bm{r'}) \Psi_0(\bm{r}') \chi(\bm{r})=
\nonumber \\
& & D_{*}\int d\bm{r'}
 \Psi_0(\bm{r'})V_{D}(\bm{r}-\bm{r'})\Psi_0(\bm{r}') \chi(\bm{r})+
 s \Psi_0^2(\bm{r}) \chi(\bm{r}),  \nonumber \\
\lefteqn{(X \chi)(\bm{r}) = } \label{eq:CX} \\
& & (N-1)\int d\bm{r'} \Psi_0(\bm{r'}) V(\bm{r}-\bm{r'})
\chi(\bm{r'}) \Psi_0(\bm{r})= \nonumber \\
& & D_{*} \int d\bm{r'} \Psi_0(\bm{r'}) V_{D}(\bm{r}-\bm{r'})
 \chi(\bm{r'}) \Psi_0(\bm{r})+ s \Psi_0^{2}(\bm{r}) \chi(\bm{r}),
\nonumber 
\end{eqnarray}
for $\chi=u,\upsilon$. The $C$ operator describes the usual direct
interaction, while the $X$ operator describes exchange interaction
between an excited quasi-particle and the condensate. Note that, in the
same notation, the stationary state $\Psi_0$ satisfies
$(H_0-\mu+C)\Psi_0=0$. In Eqs. (\ref{eq:CX}) we
have chosen the phase of $\Psi_0$ so that it is real valued.

By making the change of variables $u=\frac{1}{2}\left(f-g\right)$ and
 $\upsilon=\frac{1}{2}\left(f+g \right)$,
Eq.~(\ref{eq:mat1}) is transformed into the more convenient form 
\begin{eqnarray}
\left( \begin{array}{cc}
0 & H_0-\mu+C \\
H_0-\mu+C+2 X & 0\\
\end{array} \right) \left( \begin{array}{c}
f \\ g \end{array} \right)=
\omega
\left( \begin{array}{c} f\\ g \end{array}\right).
\label{eq:mat2}
\end{eqnarray}
The eigenvalues $\omega$ come in pairs: if $\omega$ is an eigenvalue of
Eq.~(\ref{eq:mat2}) with eigenfunction $(f,g)$, then $-\omega$ is an
eigenvalue with eigenfunction$(f,-g)$. This originates in the symmetry
of Eq.~(\ref{eq:form}) under the 
exchange of $u$ and $\upsilon^*$ with $\omega\rightarrow -\omega$.

Taking the square of the matrix
in Eq.~(\ref{eq:mat2}) gives a block
diagonal matrix, and as a result we obtain the two separate problems \cite{Huepe03}:

\begin{subequations}
\begin{eqnarray}
(H_0-\mu+C)(H_0-\mu+C+2X)f=\omega^2 f,
\label{eq:block:a}
\end{eqnarray}
\begin{eqnarray}
(H_0-\mu+C+2X)(H_0-\mu+C)g=\omega^2 g.
\label{eq:block:b}
\end{eqnarray}
\label{eq:block}
\end{subequations}
For finding the eigenvalues $\omega^2$, it is sufficient to solve one
of these equations. If one solves the the equation for $f$ (or $g$),
then the corresponding solution $g$ (or $f$) for the same $\omega$ can
be obtained via Eq.~(\ref{eq:mat2}), provided $\omega \neq 0$. Note
that $g=\Psi_0$ is a solution of Eq.~(\ref{eq:block:b}) with
$\omega$=0. This neutral mode is due to the arbitrariness in fixing
the phase of $\Psi_0$. Eq.~(\ref{eq:block:a}) also has a
neutral mode \cite{Huepe03}. For a stable ground state, all
eigenvalues $\omega$ are real and the excitation energy of one
particle into a given mode is given by (the positive signed)
$\omega$. In this case the functions $u$ and $\upsilon$ are real
valued. The appearance of negative $\omega^2$ solutions of
Eqs.~(\ref{eq:block}) , i.e. complex $\omega$, indicates instability of
the condensate.

In our application we find the eigenvalues of Eq.~(\ref{eq:block:a})
by first discretizing $f$ on a 2D grid, as described for the ground
state $\Psi_0$ in section~\ref{sec:ground}. The eigenstates may be
classified as odd or even with respect to reflection through the x-y
plane. Thus, as in the case of the ground state, only the positive $z$
semi-axis need to be sampled. The calculation of the integrals with
$V_D(\bm{r})$ in Eqs.~(\ref{eq:CX}) is performed in momentum space via
the use of Eq.~(\ref{dipintegral}) (with the appropriate
re-interpretation of $n(\bm{r'})$ there) and the DHFT. The excitation modes
with $m>0$ require special attention, and we refer the reader to the
appendix for their treatment. As part of the DHFT we need the Fourier
transform in the $z$ direction. This is performed by a fast cosine
transform for even parity, and fast sine transform for odd parity
\cite{NumericalRecipes}, for which we used the FFTW software package.

Since (as we find) many excited modes may be obtained to very high
accuracy with relatively small grids (32x32 to 64x64), it is feasible
to construct the matrix elements of $A \equiv
(H_0-\mu+C)(H_0-\mu+C+2X)$ and diagonalize it. Since we are only
interested in the lowest energy eigenstates, the Arnoldi method
\cite{Arnoldi51} is a much more efficient method, which is also
applicable to much larger, 3D grids \cite{Huepe03}. It is an
iterative method that requires at each iteration only the action of
$A$ on the vector $f$, and the full matrix $A$ itself need not be
constructed. It is particularly effective for sparse matrices. In our
case, $A$ is not sparse, but it has a special structure: parts of it
are diagonal in space and other parts (the kinetic energy as well as
the dipolar parts of $C$ and $X$) are diagonal in momentum space. The
use of DHFT enables the efficient calculation of its action on $f$
without ever forming the full matrix of $A$ in one particular basis.
Interestingly, this is the same property that makes the time-dependent
propagation technique \cite{Ruprecht96} appealing for calculating the
spectrum of excitations: in that case, only successive operations with
the Hamiltonian on the wavefunction are necessary. However, combined
use of linearization (i.e BdG equations) and the Arnoldi method should
be much more efficient, even in the 3D case. Of course, non-linear
effects, which in principle can be probed by the time-dependent
method, cannot be studied by use of the BdG equations alone. More
implementation details are given in the appendix.

For completeness of the discussion, we compare in the next subsection
the exact numerical solution of the BdG equations as obtained with our
new method, with some low lying modes computed with the time-dependent
variational method \cite{PerezGarcia96,Yi01,Yi02,Goral02}. In the
variational method, one assumes a time-dependent Gaussian ansatz:
\begin{eqnarray}
\psi(\bm{r},t)=A(t)\prod_{\eta=x,y,z}e^{[\eta-\eta_{0}(t)]^2/2w^2_{\eta}+
i\eta\alpha_{\eta}(t)+i\eta^2\beta_{\eta}(t)},
\label{eq:var}
\end{eqnarray} 
with the parameters $A$ (complex amplitude), $w_{\eta}$ (width),
 $\eta_{0}$ (center of cloud), $\alpha_{\eta}$ and $\beta_{\eta}$ are
 variational parameters. The resulting equations of motion give the
 equilibrium widths (variational Gaussian solution of the
 time-independent GP equation) and frequencies of some low lying
 modes. The modes that are described by the ansatz of
 Eq.~(\ref{eq:var}) are, first, the three ``sloshing'' or Kohn modes
 corresponding to the movement of the center of the cloud
 $\eta_{0}$. These are found to have the frequencies $\omega_{\rho}$
 and $\omega_{z}$ of the harmonic oscillator and are not affected by
 the interaction. In fact, Kohn's theorem \cite{Kohn61} proves that
 the \textit{exact} solution for these modes gives the same harmonic
 oscillator frequencies, and are not affected by the interaction. To
 understand this, consider a small displacement of the center of the
 mass of the cloud without changing its shape. The inter-cloud forces
 are then unchanged, while the restoring force due to the harmonic
 trap is proportional to the displacement and is the same throughout
 the cloud. This results in a classical harmonic motion of the cloud
 as a whole. The constant frequency of these modes will provide a
 good check on the numerical accuracy of our algorithm. Secondly,
 one obtains three collective modes describing the oscillation of the
 widths, two modes with $m=0$ and one with $m=2$. In the ideal gas
 limit they correspond to $m=0$ modes with frequencies $2
 \omega_{\rho}$ and $2\omega_{z}$, and $m=2$ mode with frequency
 $2\omega_{\rho}$. These modes have been illustrated graphically in
 \cite{Yi01,Yi02,Goral02}, with the two $m=0$ modes described as the
 breathing mode and the quadrupole mode.

\subsection{Behavior and shape of BdG excitations\label{sec:BdGresults}}

The parameter space of the dipolar BEC problem in a cylindrical trap
is 3 dimensional: we have the aspect ratio of the trap,
$\frac{\omega_{z}}{\omega_{\rho}}$, the dipolar parameter $D_{*}$, and
the contact interaction parameter $s$. In this work we shall
concentrate on dominant dipole-dipole interactions, and set the
contact interaction to zero. We explore the behavior of the BdG modes
with varying dipole-dipole interaction strength in different trap
geometries, from pancake shaped to cigar shaped. For orientation,
consider a $^{52}$Cr gas \cite{Stuhler05} with magnetic dipole moment
$6\mu_B$. Assume it is confined in a trap with
$\omega_{\rho}=2\pi\times 200$Hz. Then $D_{*}=0.0024(N-1)$. E.g, for
$N=1000$ atoms, $D_{*}=2.4$. We assume that it would be possible, through a
Feshbach resonance, to make the scattering length zero \cite{Werner05,Pavlovic05}.

\begin{figure}
\resizebox{3.4in}{!}{\includegraphics{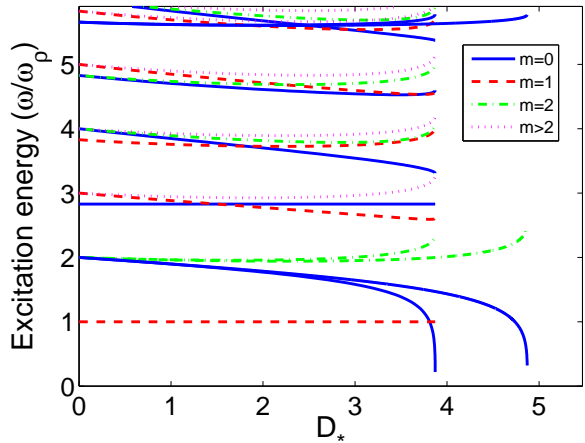}}
\caption{Excitations frequencies as function of the dipole parameter
$D_{*}$ for dipolar BEC in the JILA pancake trap
($\omega_{z}/\omega_{\rho}=\sqrt{8}$), with zero scattering
length. Plotted are modes with $m=0-4$. The 3 lines that extend to
higher $D_{*}$ are the variational results (cf. Fig.~2 of Ref.~\cite{Goral02}).
\label{fig:JILApan1}}
\end{figure}

Let us first consider a BEC in a JILA pancake trap \cite{Jin96} with
$\frac{\omega_{z}}{\omega_{\rho}}=\sqrt{8}$. The results are presented
in Fig.~\ref{fig:JILApan1}. For $D_{*}=0$ we retrieve the ideal-gas
results $\frac{\omega}{\omega_{\rho}}=n_{z}\sqrt{8}+n_{\rho}$ with
$n_{z},n_{\rho}=0,1,2,\ldots$. The lowest mode has $m=1$ and
corresponds to a transverse Kohn mode with frequency $\omega_{\rho}$
(in fact these are two degenerate modes: m=1 and m=-1, or
alternatively, sloshing motion in the $x$ and $y$ directions). The
frequency is evidently constant as a function of $D_{*}$, in agreement
with Kohn's theorem. Similarly, the second lowest $m=0$ mode is a Kohn
mode in the $z$ direction with frequency $\omega_z$. Next, consider
the two modes that converge to $\omega=2\omega_{\rho}$ in the
ideal-gas limit. One of them has $m=0$ (solid line) and the second
$m=2$ (dashed line). The $m=0$ mode is shifted down in frequency with
increasing $D_{*}$ while the $m=2$ mode is shifted up. The $m=0$ mode
goes to zero at $D_{*}=3.87$. This point marks the collapse of the
condensate: for higher value of $D_{*}$, there is no stable solution
of the GPE. These two modes are also described by the variational
method outlined above, with good agreement with the exact numerical results up to
about $D_{*}=2.2$. However, the variational method significantly
overestimates the $D_{*}$ for collapse (giving $D_{*}=4.87$ at
collapse). Our exact numerical results are in agreement with the
numerical results for these two modes obtained in Ref.~\cite{Goral02} using
the time-dependent response of the system to external perturbation,
and moreover, we are able to resolve them right down to the collapse
point. 

The behavior of the next few low modes is also interesting: some modes
of different symmetries cross each other, while two modes (one of the
them the third $m=0$ mode) converge together near to collapse. Note
also the 5'th $m=0$ mode, with the ideal-gas frequency of
$2\omega_z$. This mode is also described by the variational method. We
see that the lowest $m=0$ variational mode is also the lowest $m=0$
exact mode, but the second variational $m=0$ mode is much higher, and
between them there are two $m=0$ modes (excluding the Kohn mode), as
well as other $m>0$ modes, which are not described at all by the
variational method. As mentioned above, the variational method obtains
only two $m=0$ modes (excluding a Kohn mode) that approach
$2\omega_{\rho}$ and $2\omega_{z}$ in the ideal gas limit. But for
$\omega_{z}>2 \omega_{\rho}$ there are at least two other (non-Kohn)
$m=0$ modes that are between them, and approach
$2\omega_{\rho}+\omega_{z}$ and $4\omega_{\rho}$ in the ideal gas
limit.

It is worth noting the computational efficiency of our algorithm in
obtaining these BdG results. We obtained 100 converged modes for a
given dipole moment and $m$ in about half a minute on our PC, with a
grid size of $N_{\rho}$x$N_z=$34x64.

\begin{figure}
\resizebox{3.4in}{!}{\includegraphics{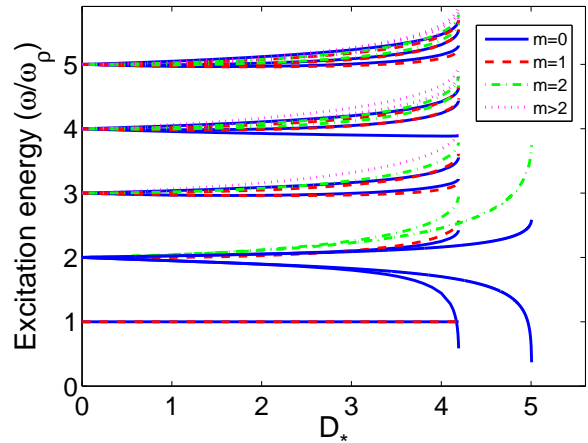}}
\caption{Excitations frequencies as function of the dipole parameter
$D_{*}$ for dipolar BEC in a spherical trap with zero scattering
length. Plotted are modes with $m=0-4$. The 3 lines that extend to
higher $D_{*}$ are the variational results.
\label{fig:spherical}}
\end{figure}

Let us now consider the case of a spherical trap
$\omega_{\rho}=\omega_z \equiv \omega_0$, Fig.~\ref{fig:spherical}.
The lowest excitation (in the ideal gas limit) consists of three
degenerate Kohn modes corresponding to $l=1$ with $m=0,-1,+1$. Note
that the Kohn frequencies are constant and maintain their degeneracy
for $D_{*}>0$, even though the dipolar interaction does not conserve
the total angular momentum, and the ground state shape is elongated in
the $z$ direction. Next, consider the 4 modes that converge to
$2\omega_0$: two with $m=0$, and another two with $m=1$ and
$m=2$. Actually, there are six, accounting for the degeneracy with
negative $m$. In the ideal gas limit these correspond to five
degenerate $l=2$ modes and one $l=0$ mode. The lowest $m=0$ of these
causes the collapse of the condensate at $D_{*}=4.19$. Note that this
mode crosses the three Kohn modes. This is possible whenever there are
two different $m$ modes or two modes with the same $m$ but different
parity with respect to the symmetry $z \rightarrow -z$. The two $m=0$
modes and the $m=2$ mode are also described by the variational
method. The variational frequencies agree quite well with the exact
ones up to $D_{*} \approx 3$, although, again, the variational method
overestimates the critical $D_{*}$ for collapse.

Note that an isotropic short range interaction that conserves $l$
would only split the degenerate ideal gas levels into different $l$
modes. Thus, if there is a dominant short range interaction and a
smaller dipolar interaction in a spherical trap, the dipolar
interaction would not just shift the levels, but also split them into
non-degenerate states of different $|m|$. This could provide an
interesting and unambiguous experimental signature of dipolar
interaction effects. In such an experiment, it would be important to
verify the spherical harmonicity of the trap, to exclude splitting due
to anisotropy of the trap or non-harmonicity. This could be
accomplished by measuring the Kohn (i.e, sloshing, or dipole) modes.

\begin{figure}
\resizebox{3.4in}{!}{\includegraphics{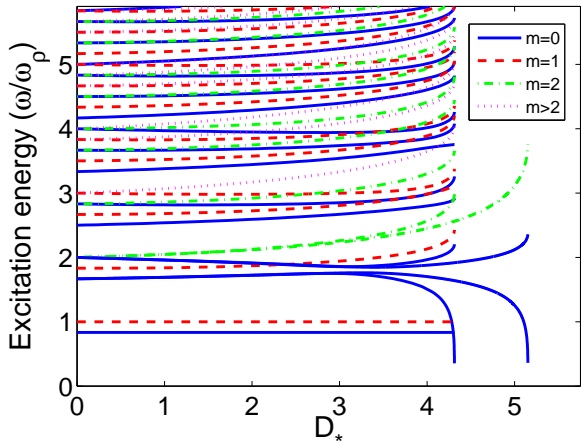}}
\caption{Excitations frequencies as function of the dipole parameter
$D_{*}$ for dipolar BEC in a cigar trap with
$\omega_{z}/\omega_{\rho}=5/6$ and with zero scattering
length. Plotted are modes with $m=0-4$. The 3 lines that extend to
higher $D_{*}$ are the variational results.
\label{fig:cigar12}}
\end{figure}

We move now to a slightly cigar shaped trap, with
$\omega_{z}/\omega_{\rho}=5/6$ (Fig.~\ref{fig:cigar12}); Here, the
collapse occurs at $D_{*}=4.32$. The interesting feature here is the
avoided crossing between the second and third $m=0$ modes. We find
that in the avoided crossing the nature of the lowest mode changes from
quadrupole-like mode for small $D_{*}$ to a breathing mode close to
collapse. However, we found the same change in the nature of the lowest
mode also in a spherical trap, where there is no such avoided crossings
(see below. See also \cite{Goral02} for discussion of the nature of
the modes within the variational method
\footnote{Note a misprint there, switching ``breathing'' with
``quadrupole-like'' in the discussion for the aspect ratio range
$0.75<l<1.29$ ($l\equiv (\omega_{\rho}/\omega_z)^{1/2}$).}).

\begin{figure}
\resizebox{3.4in}{!}{\includegraphics{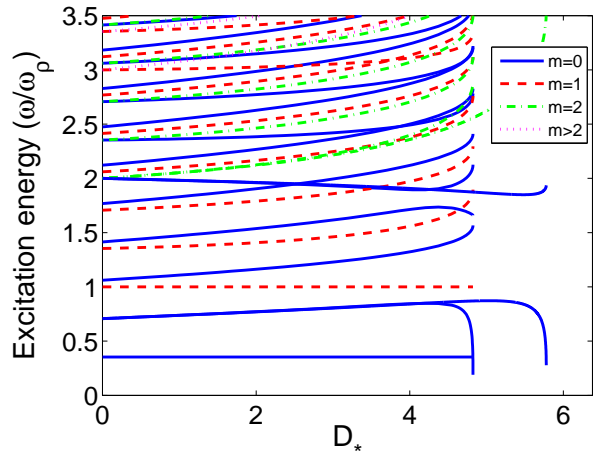}}
\caption{Excitations frequencies as function of the dipole parameter
$D_{*}$ for dipolar BEC in a JILA cigar trap
($\omega_{z}/\omega_{\rho}=1/\sqrt{8}$) with zero scattering
length. Plotted are modes with $m=0-4$. The 3 lines that extend to
higher $D_{*}$ are the variational results (cf.~Fig.~3 of Ref.~\cite{Goral02}).
\label{fig:JILAcig1}}
\end{figure}

Finally, consider the JILA cigar trap \cite{Jin96} with
$\omega_{z}/\omega_{\rho}=1/\sqrt{8}$, Fig.~\ref{fig:JILAcig1}. Here
collapse occurs at $D_{*}=4.83$. The two $m=0$ modes which are
described by the variational method have now a wide gap in energy and
there is no longer a clear avoided crossing between them. There is an
interesting pattern of crossing between some of the higher modes. Note
again, that the lowest variational mode is the one that leads to
collapse, but the two other two variational modes lie above others
which are not accounted for by the variational ansatz.

\begin{figure}
\resizebox{3.4in}{!}{\includegraphics{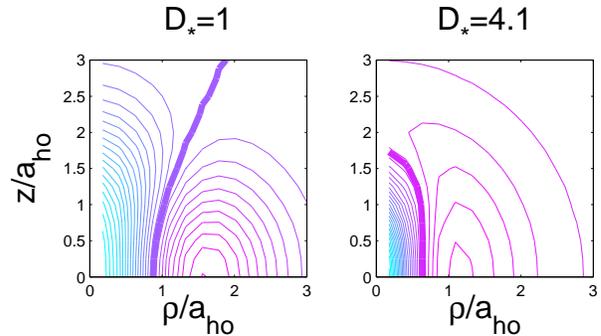}}
\caption{The spatial density perturbation $f=u+v$ (see text) for 
the lowest (non Kohn) mode in a spherical trap. Left: for $D_{*}=1.0$,
right: for $D_{*}=4.1$, close to collapse. The nodal line is the
thick line.
\label{fig:waves1}}
\end{figure}
We can also examine the shape and nature of the BdG eigenmodes. The
two eigenmode functions $u$ and $\upsilon$ are real, and determine the
time dependent oscillation through Eqs.~(\ref{eq:perturb}) and
(\ref{eq:form}). The oscillation of the density
$\rho(\bm{r},t)=|\psi(\bm{r},t)|^2$ is then given, to linear order in
$u,\upsilon$, by $\delta\rho(\bm{r},t)=2
(u(\bm{r})+\upsilon(\bm{r}))\cos(\omega t)$. The function
$f=u+\upsilon$ therefore gives the shape of the density oscillations.

As an example, consider the lowest (non Kohn) mode in the spherical
trap, the mode that goes to zero in Fig.~\ref{fig:spherical} and leads
to the collapse of the BEC. In Fig.~\ref{fig:waves1} we draw the
contour plot of $f$ for this mode, for two values of $D_{*}$, one
small, and one close to the collapse point. If we examine the nodal
line ($f(\bm{r})=0$, heavy line), we see that for small $D_{*}$ it forms an open,
hyperbolic-like contour. This signifies a quadrupole-like mode. In
contrast, close to collapse, the nodal line forms a closed, elliptic
contour, typical of a breathing mode. We conclude that the nature of
the mode changes from quadrupole-like for small $D_{*}$ to a breathing
mode for $D_{*}$ close to collapse. This conclusion is in agreement
with analysis based on the variational method \cite{Goral02}. In
between, the nodal line is parallel to the $z$ axis, and the eigenmode
character is essentially that of a pure transverse ($\rho$)
excitation.

\subsection{Collective versus single-particle excitations}

\begin{figure}
\resizebox{3.4in}{!}{\includegraphics{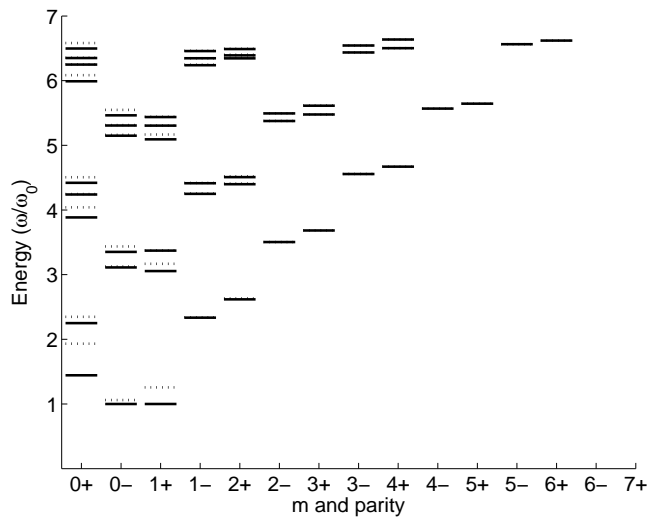}}
\caption{Excitation spectrum in a spherical trap with $D_{*}=4$. The
eigenenergies of BdG equations (\ref{eq:mat1}) are represented by
thick solid bars. Dotted bars correspond to the single-particle
spectrum of the HF Hamiltonian (\ref{eq:sp}).
\label{fig:modes}}
\end{figure}

Let us now examine in more detail the structure of the BdG excitations
spectrum for a specific case. In Fig.~\ref{fig:modes} we show the
spectrum evaluated for a spherical trap with $D_{*}=4$. Each state is
characterized by angular momentum projection $m$, and by even
(positive) parity or odd (negative) parity with respect to reflection
$z \rightarrow -z$. The BdG states are represented by thick solid
bars.

The BdG modes in each column of Fig.~\ref{fig:modes} are grouped in
multiplets with increasing near degeneracy of $1,2,3,\ldots$ near the
harmonic oscillator frequencies. In the ideal-gas limit these groups
become exactly degenerate. Note that the for the $0^+$ column the two
lowest modes (which are far apart) become degenerate (with frequency
$2\omega_0$) at the ideal gas limit. The splitting between the groups
in each given column is approximately $2\omega_0$. The splitting
between the towers of even and odd modes of the same $m$ is
approximately $\omega_0$. In the ideal gas limit, the tower of states
with $m^+$ is degenerate with that of $(m-1)^-$. We can classify the
states in the ideal gas limit by ($l,n_r$), with $l$ the total angular
momentum, $n_r$ the number of radial nodes (not counting a node at
$r=0$), and with energy $(2n_r+l)\hbar\omega_0$. The towers of $m^+$
and $(m-1)^-$ states become, in the ideal gas limit, a tower of
$(l,n_r)$ states as follows: the lowest state is
$(l,n_r)=(m,0)$. Above it there is a pair $(m+2,0),(m,2)$ degenerate
in the ideal gas limit, followed by a triplet $(m+4,0),(m+2,2),(m,4)$ and so
on. Here, the ordering of the states inside each multiplet is conventional only and
does not indicate their order of increasing energy when split by the
interaction.

The BdG eigenmodes are given by the pair $u,\upsilon$ corresponding to
positive and negative frequencies. A collective mode is characterized
by non negligible $\upsilon$ component. It describes excitation of a
quasi-particle, as opposed to an excitation of a single
particle. However, the high-energy part of the spectrum is expected to
be well reproduced by a single-particle description in the mean-field
approximation \cite{Dalfovo97,BEC2003}, since the condensate as a
whole has little time to respond to the fast oscillations of a single
particle in a high-frequency mode. The single-particle, or
Hartree-Fock (HF) picture is
obtained by neglecting the coupling between $u$ and $\upsilon$ in
Eq.~(\ref{eq:BdGpair}). This corresponds to setting $\upsilon=0$ in
the first equation of Eq.~(\ref{eq:BdGpair}), which then reduces to
the eigenvalue problem $(H_{HF}-\mu)=\omega u$, with the
HF Hamiltonian
\begin{eqnarray}
H_{HF}=H_0+C+X.
\label{eq:sp}
\end{eqnarray}
In this case, the eigenfunctions $u(\bm{r})$ satisfy the normalization
condition $\int u_i^{*}(\bm{r})u_j(\bm{r})=\delta_{ij}$. 

The spectrum
of $H_{HF}$ is shown as dotted horizontal bars in
Fig.~\ref{fig:modes}. The general structure of the HF spectrum is very
similar to that obtained with the BdG equations (\ref{eq:BdGpair}),
apart from states with low energy and $m$, which are collective modes.
Note that the HF spectrum fails to satisfy the Kohn
theorem for the dipole modes. It is worth noting the good agreement
between the HF and the BdG spectra for $m>2$ as well as
for the odd $m=1$ modes. A close look at the near-degenerate groups of
the even $m=0,1$ modes shows that for the $0^+$ symmetry, there is
good agreement between the two spectra except for two modes in each
group. For the $0^-$ and $1^+$ symmetries, there is good agreement
except for one mode in each group. These observations can be
understood as follows. In the ideal gas limit, modes with $l>0$, have
zero amplitude at the center of the condensate, where the density is
at its maximum. With increasing $l$ the modes become more concentrated
near the surface. The dipolar non-diagonal coupling term in
Eq.~(\ref{eq:mat1}) is proportional to the local ground state amplitude,
and so becomes small for surface modes. This, assuming that modes
having $l>1$ in the ideal gas limit are well approximated by the
HF description, explains the pattern of
Fig.~\ref{fig:modes}, except for the even $m=0$ tower.

To explain the fact that for the $0^+$ modes we see two non
single-particle modes in each near-degenerate group, we observe that,
to first order of perturbation theory, the dipolar interaction mixes
the $l=0$ and $l=2$ modes into two orthogonal linear combinations,
each having some $l=0$ component. Thus, for example, the group of
three $0^+$ near degenerate modes around $\omega/\omega_0 \approx 4$
corresponds to the ideal gas modes $(l,n_r)=(0,2),(2,1),(4,0)$. The
first two are mixed already in first order of perturbation theory, so
that both have some $l=0$ character, and as a consequence are not well
described by the single-particle picture. This interpretation is
supported by visual examination of the exact numerical eigenfunctions
of the three states. Note that the state that goes to the $(4,0)$
state in the ideal gas limit is the second of the three in order of
energy.

\subsection{Quantum depletion}

The quantum depletion, i.e, the number of particles $\tilde{N}$ out of
the condensate, due to the interaction, is given by the Bogoliubov
theory as:
\begin{eqnarray}
\tilde{N}=\int d \bm{r}\tilde{n}(\bm{r}),
\label{eq:deptot}
\end{eqnarray}
with the local depletion defined by
\begin{eqnarray}
\tilde{n}(\bm{r})=\sum_{j} |\upsilon_{j}(\bm{r})|^2,
\label{eq:dep}
\end{eqnarray}
where the ``hole'' components $\upsilon_{j}$ are obtained by solving
the BdG equations (\ref{eq:mat1}).

Typically, many thousands of modes need to be obtained in order to
converge the sum in Eq.~(\ref{eq:dep}) \cite{Dalfovo97}. A useful
approximation in this context in the local density approximation (LDA)
\cite{Dalfovo97,Reidl99}, which is the leading order of a
semi-classical approximation. It was employed for the description of
BEC with a repulsive short range interaction. For an attractive short range
interaction, or more generally, for a dipolar BEC with $d^2>a/3$
\cite{Eberlein05}, an homogeneous BEC is unstable to small momentum
perturbations. Thus, the LDA, which assumes local homogeneity, leads to
unphysical complex frequencies for small momenta. However, it is still
useful to describe the high momentum modes.

The LDA amounts to setting
\begin{eqnarray}
u_{j}(\bm{r})\rightarrow u(\bm{p},\bm{r})e^{i \bm{p \cdot
  r}},\nonumber \\
\upsilon_{j}(\bm{r})\rightarrow \upsilon(\bm{p},\bm{r})e^{i \bm{p \cdot
  r}},\nonumber \\
\sum_{j} \cdots \rightarrow \int \frac{d^3\bm{p}}{(2\pi)^3}\cdots,
\end{eqnarray}
where $u(\bm{p},\bm{r}),\upsilon(\bm{p},\bm{r})$ are normalized by
$|u(\bm{p},\bm{r})|^2-|\upsilon(\bm{p},\bm{r})|^2=1$. In the
semi-classical limit the functions $u(\bm{p},\bm{r})$ and
$\upsilon(\bm{p},\bm{r})$ are slowly varying on the scale of the trap
size, hence their derivatives are negligible. We then obtain the same
structure of the BdG equations~(\ref{eq:mat1}), with the operators
$H_0, X$ of Eqs.~(\ref{eq:CX}) replaced by their LDA versions:
\begin{eqnarray}
H_0^{ld}&=&\frac{p^2}{2m}+U(\bm{r}), \nonumber \\
(X^{ld} \chi)(\bm{p},\bm{r})& =& (D_{*} \tilde{V}_{D}(\bm{p})+ s)
\Psi_0^{2}(\bm{r}) \chi(\bm{p},\bm{r}),
\end{eqnarray}
where $\tilde{V}_{D}(\bm{p})$ is given by Eq.~(\ref{dippot}). The
operator $C$ of Eq.~(\ref{eq:CX}) remains
unchanged, while the exchange operator $X^{ld}$ becomes local. Thus,
in the LDA all the operators are local, and the solution of the BdG
equations becomes algebraic. The excitation frequency is given by:
\begin{eqnarray}
\omega(\bm{p},\bm{r})=\sqrt{\omega_{HF}^2(\bm{p},\bm{r})-X^2(\bm{p},\bm{r})},
\label{eq:LDAfreq}
\end{eqnarray}
where $\omega_{HF}$ is the HF frequency within the LDA, given by
\begin{eqnarray}
\omega_{HF}(\bm{p},\bm{r})=H_0^{ld}(\bm{p},\bm{r})-\mu+C(\bm{r})+X^{ld}(\bm{p},\bm{r}).
\end{eqnarray}

In the semi-classical approximation, one replaces the sum over the
discrete states in Eq.~(\ref{eq:dep}) with integral over
\begin{eqnarray}
\upsilon^2(\bm{p},\bm{r})=\frac{\omega_{HF}(\bm{p},\bm{r})-\omega(\bm{p},\bm{r})}{2\omega(\bm{p},\bm{r})}.
\end{eqnarray}

Since the LDA is inappropriate for the low lying modes, we may
calculate the contribution to Eq.~(\ref{eq:dep}) from exact, discrete
BdG modes up to a certain frequency cutoff $\omega_c$, and use the LDA to obtain the
contribution from higher frequency modes. Then Eq.~(\ref{eq:dep}) is
replaced by
\begin{eqnarray}
\tilde{n}(\bm{r})=\sum_{j}
|\upsilon_{j}(\bm{r})|^2\Theta(\omega_{c}-\omega_{j})+
\int_{\omega_c}^{\infty}d\omega \tilde{n}(\omega,\bm{r}),
\end{eqnarray}

with 
\begin{eqnarray}
\tilde{n}(\omega,\bm{r})=
\int \frac{d\bm{p}}{(2\pi)^3}
\upsilon^2(\bm{p},\bm{r}) \delta(\omega(\bm{p},\bm{r})-\omega) \times
\nonumber \\
\Theta(\omega_{HF}(\bm{p},\bm{r})).
\label{eq:dep2}
\end{eqnarray}
Note that the factor in the Dirac delta function depends on the
direction in momentum space. That is, the iso-energy surfaces are not
surfaces of equal momentum $p$. Also, we use the Heaviside function
$\Theta(\omega_{HF}(\bm{p},\bm{r}))$ to exclude unphysical LDA modes with
$\omega_{HF}<0$, even though the definition (\ref{eq:LDAfreq}) can assign
them a real positive frequency, due to taking the positive root. 
However, this matters only for small momenta below $\omega_c$.

As an example, we consider the case of a BEC in a spherical trap with
$D_{*}$=4.0 and $s=0$. Using Eqs.~(\ref{eq:dep2}) and
(\ref{eq:deptot}), we find that the total depletion is 1.4
particles. The fractional depletion depends on the number of
condensate particles. Since $D_{*}$ is proportional to $(N-1)d^2$, we
can achieve the same value of $D_{*}$ with many particles with a small
dipole moment, or a few particles with a large dipole moment. For the
$^{52}$Cr example mentioned in the beginning of
section~\ref{sec:BdGresults}, we obtain $D_{*}=4.0$ with $N=1670$, and
the quantum depletion is entirely negligible. However, we may imagine
a dozen or so molecules with high dipole moment in a micro-trap. In
which, case, the depletion may be measurable.

To demonstrate the agreement between the LDA and the exact BdG
spectrum for high frequencies, we define the spectral distribution
$g_{ld}(\omega)$ as the number of depleted bosons per unit
frequency. Thus $g_{ld}(\omega) d\omega$ is the number of depleted
bosons with frequency $\omega$ in the interval $d\omega$. We have
\begin{eqnarray}
g_{ld}(\omega)=\int d\bm{r} \tilde{n}(\omega,\bm{r}).
\label{eq:dislda}
\end{eqnarray}

We want to compare it with the spectral distribution $g$:
\begin{eqnarray}
g(\omega)=\sum_{j}\int d\bm{r}
\tilde{n}_j(\bm{r})\frac{1}{2 \sqrt{\pi\sigma^2}}\exp
(-\frac{1}{2}(\frac{\omega-\omega_j}{\sigma})^2),
\label{eq:dis}
\end{eqnarray}
obtained from the low-lying discrete modes by folding
$\tilde{n}_j(\bm{r})$
with a Gaussian of standard deviation $\sigma$ (some smoothing of the
discrete data is is necessary for meaningful comparison with the
continuous LDA spectral distribution).

\begin{figure}
\resizebox{3.4in}{!}{\includegraphics{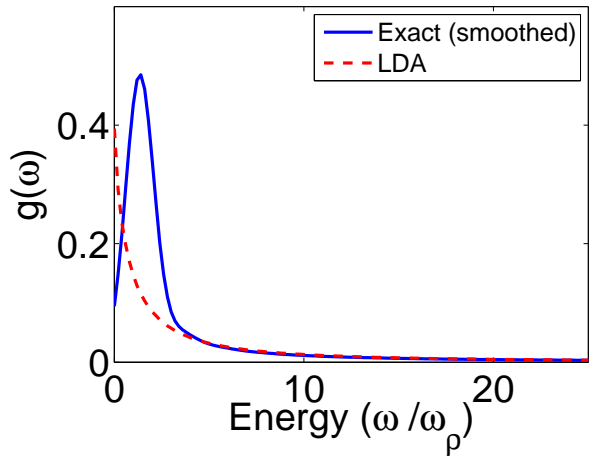}}
\caption{The spectral distribution $g_{ld}(\omega)$ in the local
 density approximation, Eq.~(\ref{eq:dislda}), is compared with the
 smoothed, numerically exact spectral distribution $g_(\omega)$,
 Eq.~(\ref{eq:dis}), for BEC in a spherical trap with $D_{*}=4.0$.
\label{fig:dep}}
\end{figure}

In Fig.~\ref{fig:dep} we compare $g$ and $g_{ld}$ for a BEC in a
spherical trap, with dipolar interaction $D_{*}=4.0$ and $s=0$. We
computed $g$ from Eq.~(\ref{eq:dis}) with $\sigma=\omega_{0}$. For frequencies
above $\omega/\omega_0 \approx 5$ the two curves are essentially
indistinguishable. This shows the efficiency of the LDA in describing
the high energy part of the spectrum.

\section{Conclusions}

We have calculated, for the first time, many BdG excited modes of a
dipolar BEC in a cylindrical trap, by direct solution of of the BdG
equations. This achievement was made possible by a novel, highly
efficient algorithm which takes advantage of the cylindrical
symmetry. We showed how properties of the BdG spectrum depend on the
shape of the trap; examined in detail the spectrum in a spherical trap
and the nature of the modes (collective vs single-particle); and
calculated the quantum depletion due to dominant dipolar interactions,
which is typically very small, but may become more significant in
micro-traps containing a dozen or so molecules with high dipole
moment.

We note that the formalism developed in this work may be easily
extended to compute properties of dipolar BEC and its depletion at
non-zero temperature using the Hartree-Fock-Bogoliubov-Popov method
\cite{Reidl99}. We intend to study the non-zero temperature
behavior in a future work.

\begin{acknowledgments}
SR gratefully acknowledges financial support from an anonymous fund,
and from the United States-Israel Educational Foundation (Fulbright
Program), DCEB and JLB from the DOE and the Keck Foundation.
\end{acknowledgments}

\appendix*
\section{}

\subsection{Modification of the dipolar potential\label{App:modpot}} 

In this part of the appendix we analyze the numerical accuracy of the 3D FFT
method for calculating the dipolar interaction. A correction is
suggested to increase the accuracy, and this correction also applies
to our 2D algorithm.

The 3D FFT method was used to calculate the mean field potential due
to dipolar interactions via Eq.~(\ref{dipintegral}). To check its
accuracy it is more convenient to consider the dipolar interaction
energy for a dipole strength $D_{*}=1$, given by:
\begin{eqnarray}
E_{D}= \frac{1}{2} \int\int d\bm{r} d\bm{r'} V_{D}(\bm{r}-\bm{r'}) n(\bm{r'})n(\bm{r}).
\label{Edipr}
\end{eqnarray}
Given $n(\bm{r})$, this expression can be evaluated numerically on a
3D grid by first performing the $\bm{r'}$ integration using
Eq.~(\ref{dipintegral}), which requires FFT and inverse FFT, and then
performing the $\bm{r}$ integration as discrete summation on the
spatial grid. Note that $\tilde{V}_D(k)$ in Eq.~(\ref{dipintegral}) is given
analytically by Eq.~(\ref{dippot}) and does not require FFT.

Eq.~(\ref{Edipr}) may be alternatively written in the form:
\begin{eqnarray}
E_{D}= \frac{1}{2(2\pi^3)} \int d\bm{k} \tilde{V}_{D}(k) \tilde{n}(\bm{k})^2.
\label{Edipk}
\end{eqnarray}
This expression may be evaluated numerically by using one FFT to
obtain $\tilde{n}(k)$ and then performing the integration as summation
in momentum space. The two numerical procedures give the same result
up to machine precision.

For a Gaussian density
\begin{eqnarray}
n(\bm{r})=\frac{1}{\pi^{3/2}\sigma^2\sigma_z}\exp(- (x^2+y^2)/\sigma^2-z^2/\sigma_z^2) 
\end{eqnarray}
it is possible to obtain an analytic expression for $E_{D}$ using
Eq.~(\ref{Edipk}) \cite{Martikainen02,Yi00}. This enables us to check the
accuracy of the numerical calculation.

For a spherically symmetric Gaussian, $\sigma=\sigma_z$, defined on a
cubic grid with equal resolution in the three axis, we find numerically
the exact result $E_D=0$. In fact, by noticing that
$\tilde{V}_D(k)=\frac{4\pi}{3}\frac{2 k_z^2-k_x^2-k_y^2}{k^{2}}$, it
is seen that $E_{D}$ evaluated on a cubic grid is guaranteed to be zero
for any density $n(x,y,z)$ which is symmetric under all permutations of
$(x,y,z)$.

For a pancake shaped density $\sigma=2, \sigma_z=1$, the exact result
is $E_D=0.038670861\ldots$. We have evaluated it on cubic grids of
extent $[-R,R]\times[-R,R]\times[-R,R]$ with varying size $R$ and number of
points $N$x$N$x$N$ . The relative error is tabulated in the first row of
table \ref{tab:table1}. It is seen that the errors are small but far
from machine accuracy. Note that convergence with respect to $N$ is
already achieved with a modest spatial resolution of
$2R/N=2*8/32=0.5$, but there is a rather slow convergence with
increasing $R$. This make it clear that the error is not due to
failure to resolve the density function $n(r)$. For cigar shaped
densities similar behavior was observed.
\begin{table*}
\caption{\label{tab:table1}Relative error of evaluating dipolar
interaction energy using the 3D FFT method, before and after
correction, for a pancake Gaussian density (see text).}
\begin{ruledtabular}
\begin{tabular}{ccccc}
 & $R=8,N=32$ & $R=8,N=64$ & $R=16,N=64$&
 $R=16,N=128$\\
\hline
original method & 2.7E-3 & 2.7E-3 & 8.6E-5 & 8.6E-5 \\
corrected method & -1.1E-5 & -1.1E-5 & 1.8E-8 & -4.4E-14 \\
\end{tabular}
\end{ruledtabular}
\end{table*}
The reason for the behavior of the numerical error is understood if we
realize that $\tilde{V}_{D}(k)$ is discontinuous at the origin, where
$n(k)$ obtains its maximum. This discontinuity originates in the long
range and non-isotropic nature of the interaction. Thus, the numerical
accuracy converges slowly with increasing grid resolution,
proportional to $1/R$, in momentum space.

An alternative and equivalent way to understand the source of the
error is that the use of FFT implicitly assumes that we are dealing
with a 3D periodic lattice of condensates, with unit cell of size $2R$. Thus the error may be
traced down to the long range interactions between copies of
condensates in different unit cells.

An obvious correction suggests itself: since our actual condensate is
isolated and of a finite size, we can limit the range of the dipolar interaction
$V_D(r)$ such that it is the same as before for $r<R$ and and zero for
$r\geq R$. This should have no physical consequences as long as $R$ is
greater than the extent of our condensate. Then the Fourier transform
of this interaction is continuous at the origin and resolved by the
grid in momentum space. We obtain the following expression for the
corrected dipolar interaction $\tilde{V}_{D}^{cutR}(k)$:
\begin{eqnarray}
\tilde{V}_{D}^{cutR}(\bm{k})&=&
\frac{4\pi}{3} 
(1+3\frac{\cos(Rk)}{R^2 k^2}-3\frac{\sin(Rk)}{R^3 k^3}) \times \nonumber \\
& & (3 \cos^{2}\alpha-1).
\label{eq:cutR}
\end{eqnarray}

Using this corrected interaction, we obtain the much better accuracy
demonstrated in the second line of table~\ref{tab:table1}. The
remaining error depends on $R$ only due to the spatial extent of the
condensate, and fast convergence is achieved by increasing $R$ while
keeping appropriate grid resolution through increasing $N$. One may
compromise on $R$ and $N$, and still obtain at least a 100 fold
increase in accuracy as compared to using the infinite range
interaction.

For highly pancake/cigar traps the condensate has also a highly
pancake/cigar shape. In this case it is natural to work with a grid
whose extent $[-Z,Z]$ in the $z$ direction is, respectively,
smaller/larger than its extent $[-P,P]\times[-P,P]$ in the $(x,y)$
plane. Thus, fewer grid points are needed along the shorter axis. In
this case, we find that without correction, numerical errors can be
typically as large as one percent. However, truncating the interaction
outside a sphere as in Eq.~(\ref{eq:cutR}) is not very helpful in this
case, since the condition $R\leq \min(Z,P)$ must be met, which
restricts the condensate extent to less than the shorter direction.
The ideal fix would be to cut the interaction exactly by the shape of
the box, or a cylinder inscribed within it. We were unable to find
an analytic expression for the Fourier transform of a dipolar interaction
bounded by a cylinder. A partial but still helpful solution for
pancake traps is to truncate the interaction only for $|z|>Z$. We
then find:
\begin{eqnarray}
\lefteqn{V_{D}^{cutZ}(\bm{k})=} \nonumber \\
& & \frac{4\pi}{3}(3\cos^2\alpha-1)+4\pi\exp(-Z k_\rho)\Big[\sin^2\alpha\cos(Z
k_z)- \nonumber \\
& & \sin\alpha\cos\alpha\sin(Z k_z)\Big].
\label{eq:cutZ}
\end{eqnarray}
With this corrected interaction, a small $Z$ may be used as long as it
fully contains the condensate. Numerical convergence with the size $P$
will still be slow, but typically the accuracy is improved by an order
of magnitude at the least.

Finally, with our 2D algorithm combining Hankel transform in the
transverse direction and Fourier transform in the $z$ direction, we
find numerical errors of similar behavior and magnitude. In the case
of 2D, small numerical errors exists even for a spherical symmetric
density, since there is no symmetry of the grid that ensures getting
the correct zero energy, as in the 3D case. All of these errors are
significantly reduced by employing the same cutoff interactions as in
the 3D case.

\subsection{Conjugate-gradients implementation}

We describe here in some detail aspects of the conjugate-gradients
implementation specific to the problem of finding the ground state of
a dipolar condensate, see section \ref{sec:ground}.The standard
conjugate-gradients algorithm performs unconstrained minimization. In
principle a constraint could be implemented through a Lagrange
multiplier. In our implementation we rather follow the idea of
\cite{Modugno03}. To account more easily for the normalization
constraint, Eq.~(\ref{eq:constraint}), we let $\Psi\rightarrow \Psi/
\|\Psi\|$ so that the energy can be obtained for condensate wave
functions $\Psi$ with a norm different from unity. This corresponds to
dividing the terms of $E[\Psi,\Psi^*]$ in Eq.~(\ref{eq:func1}) that
are quadratic in $\Psi$ by $\|\Psi\|^2$, and the interaction term
quartic in $\Psi$ by $\|\Psi\|^4$. The modified energy functional
reads:
\begin{eqnarray}
\lefteqn{E[\Psi,\Psi^*]=} \nonumber \\
& & \int d \bm{r} \frac{\Psi^{*}(\bm{r}) H_{0} 
 \Psi(\bm{r})} {\| \Psi \|^2} + \nonumber \\
& &\frac{N-1}{2 \|\Psi\|^4} \int \int
 d\bm{r} d\bm{r'} |\Psi^{2}(\bm{r'})|V(\bm{r}-\bm{r'}) |\Psi^{2}(\bm{r})|.
\label{eq:func2}
\end{eqnarray}
This functional may now be minimized directly with no constraints. During
the minimization process it may still be numerically advantageous to normalize the
wavefunction at each step.

One ingredient of the conjugate-gradients method is a line minimization of the energy functional,
that is the minimization of 
\begin{eqnarray}
E(\Psi_0+\lambda\chi)
\label{eq:fmin}
\end{eqnarray}
with respect to $\lambda$, where $\Psi_0$ is the current trial
wavefunction and $\chi$ is the proposed direction along which to move.
An important issue for our specific problem is to find the \textit{first} minimum
encountered when moving downhill in energy along the line: whenever
$d^2>a/3$, the global minimum is a collapsed state \cite{Yi00}, while the
condensate corresponds to a stable local minimum (if such
exists). Therefore, it is important that the line minimization will
not jump to an energy valley leading to the global one, but stay in
the energy valley of the initial guess. This issue is usually not
considered as important in the textbook implementation of the
conjugate-gradients method. Following \cite{Modugno03} we use the
fact that the Eq.~(\ref{eq:fmin}) is a rational function of $\lambda$. We
then easily find the roots of $dE/d\lambda$ and the first local
minimum of $E$ encountered when one moves along the line downhill in
energy starting from $\lambda=0$. The coefficients of the numerator in the
rational function require calculation of dipolar interaction integrals
with combinations of $\Psi$ and $\chi$ such as $\int\int
d\bm{r}d\bm{r'}\Psi(\bm{r}) \chi(\bm{r})
V_D(\bm{r}-\bm{r'})\chi^{2}(\bm{r'})$, etc. These integrals can be computed
in momentum space by using DHFT and the identity:
\begin{eqnarray}
\lefteqn{\int\int
d\bm{r}d\bm{r'}n_1(\bm{r})V_D(\bm{r}-\bm{r'})n_2(\bm{r'})=} \nonumber \\
& &\frac{1}{(2\pi)^3}\int d\bm{k}\tilde{n}_1(\bm{k})\tilde{V}_D(\bm{k})\tilde{n}_2(\bm{k}).
\end{eqnarray}
After finding the local minimum along a given line we proceed with
another line minimization along a conjugate direction, and so forth
until we find a local minimum of the energy functional
Eq.~(\ref{eq:func2}). 

An additional technical ingredient in our implementation of
the conjugate-gradients algorithm is the use of pre-conditioning
\cite{Payne92}, a technique used to accelerate the convergence. Our
pre-conditioner is given in momentum space as $\frac{1}{k^2/2+M}$,
with $M=\max(E,E_k)$, $E$ is the energy given by Eq.~(\ref{eq:func2}), and $E_k$
is the kinetic energy.

\subsection{Excitations spectrum}

An important issue in calculating the $m>0$ excitations is that the
grid points given by Eq.~(\ref{rindex}) are different for different
$m$. The point is that the function $f$ of Eq.~(\ref{eq:block:a}) is
represented on the $m>0$ grid, whereas the ground state wavefunction
entering Eqs.~(\ref{eq:CX}), is defined on the $m=0$ grid. Our
solution is to interpolate the ground state wavefunction to the grid
$m>0$. The interpolation is facilitated by the fact that the roots of
the Bessel function $J_{m}(r)$ march to the right as $m$ is increased,
and those of $m$ are interlaced between those of order $m-1$.
Fortunately, a highly accurate interpolation scheme is available in
this case  \cite{Makowski82}. Similarly to the exact integration formula
(\ref{integral2}), one can derive an \textit{exact} interpolation formula
for \textit{band limited} functions satisfying Eq.~(\ref{eq:condition}):

\begin{eqnarray}
f(r)=\sum_{i=1}^{\infty}\frac{2 \alpha_{0i} J_0(2\pi K r)}
{(\alpha_{0i}^2-(2 \pi K r)^2)} \frac{f(r_i)}{J_{1}(\alpha_{0i})},
\label{eq:interpolate}
\end{eqnarray}
with the grid points $r_i=\frac{\alpha_{0i}}{2 \pi K}$. This formula
may be proved by writing $f(r)$ as the Hankel transform (of order $0$)
of $\tilde{f}(k)$. Expanding $\tilde{f}(k)$ in a Fourier-Bessel
series, the coefficients are found to be proportional to
$f(r_i)$. Evaluating the resulting expression gives
Eq.~(\ref{eq:interpolate}). This formula, exact for band limited
functions, still gives very accurate approximation for $f(r)$ such
that its Hankel transform (i.e, its 2D Fourier transform) is small for
$k>K$. It can be truncated to the first $N$ terms provided $f(r)$ is
small for $r\geq R=\alpha_{0,i+1}/K$. We used this formula to
interpolate the ground state wavefunction to the required grid points
of $m>0$. Note that this interpolation need only be done once prior to
solving the BdG eigensystem. An alternative method for calculating
$m>0$ modes using a fixed grid corresponding to $m=0$ with no
need for interpolation was suggested in Refs.~\cite{Lemoine94lett,Lemoine97}.

In our application we have taken advantage of the structure of the BdG
equations (section~\ref{sec:excitations}) to efficiently compute the
low lying spectrum. We have used a variant of the Arnoldi method (the
implicitly restarted Arnoldi method), which is implemented in the
ARPACK software package \footnote{available in Matlab through the
function \textit{eigs}}, and enables finding the $M$ largest or
smallest eigenvalues of an operator $\bm{A}$, where $M$ is selected by
the user. $\bm{A}$ need not be hermitian. For our purposes, the main
advantage of this method is that it requires as input only the
evaluation of $\bm{A}$$x$ for some vector $x$. The matrix elements of
$\bm{A}$ need not be known. The user need only provide a function that
accepts $x$ and returns $y=\bm{A}x$. The eigenvalues in the requested
part of the spectrum are then found iteratively by repeated
applications of $A$, starting from some randomly chosen $x$. In our
case, $Ax$ represents the l.h.s of Eq.~(\ref{eq:block:a}), and the
computation is facilitated, as usual, by using DHFT to move between
space and momentum space representations.

\bibliography{biblo}

\end{document}